\documentclass[10pt,letterpaper]{article}
\usepackage[top=0.85in,left=2.75in,footskip=0.75in]{geometry}

\usepackage{changepage}
\usepackage{float}

\usepackage[utf8]{inputenc}
\usepackage{graphicx}
\usepackage{url}
\usepackage{hyperref}
\usepackage{enumitem}
\usepackage{booktabs}
\usepackage{makecell}

\usepackage{textcomp,marvosym}

\usepackage{fixltx2e}

\usepackage{amsmath,amssymb}

\usepackage{cite}

\usepackage{nameref,hyperref}

\usepackage[right]{lineno}

\usepackage{microtype}
\DisableLigatures[f]{encoding = *, family = * }

\usepackage{rotating}

\raggedright
\usepackage[aboveskip=1pt,labelfont=bf,labelsep=period,justification=raggedright,singlelinecheck=off]{caption}

\usepackage[table]{xcolor}

\usepackage{tikz}
\usetikzlibrary{shapes.geometric, arrows}
\usetikzlibrary{mindmap}
\usetikzlibrary{positioning}

\makeatletter
\renewcommand{\@biblabel}[1]{\quad#1.}
\makeatother

\date{}

\usepackage{lastpage,fancyhdr,graphicx}
\usepackage{epstopdf}
\fancyheadoffset[L]{2.25in}
\fancyfootoffset[L]{2.25in}

\definecolor{light-gray}{gray}{0.95}

\newcommand{\codewhite}[1]{\colorbox{white}{\texttt{#1}}}

\begin{document}
\vspace*{0.35in}

\begin{flushleft}
{\Large
\textbf\newline{Personalized Prognostic Models for Oncology: A Machine Learning Approach}
}
\newline
\\
David Dooling\textsuperscript{1},
Angela Kim\textsuperscript{1},
Barbara McAneny\textsuperscript{1},
Jennifer Webster\textsuperscript{1}
\\
\bigskip
\bf{1} Innovative Oncology Business Solutions, Albuquerque, NM, USA
\\
\bigskip

%
%





* ddooling@innovativeobs.com

\end{flushleft}
\section*{Abstract}
We have applied a little-known data transformation to subsets of the Surveillance, 
Epidemiology, and End Results (SEER) publically available data of the National Cancer 
Institute (NCI) to make it suitable input to standard machine learning classifiers. This transformation properly treats the right-censored data in the SEER data and the resulting Random Forest and Multi-Layer Perceptron models predict full survival curves. Treating the 6, 12, and 60 months points of the resulting survival curves as 3 binary classifiers, the 18 resulting classifiers have AUC values ranging from  .765 to .885. Further evidence that the models have generalized well from the training data is provided by the extremely high levels of agreement between the random forest and neural network models predictions on the 6, 12, and 60 month binary classifiers.


\section{Introduction}
\label{sec:introduction}

Opportunities are emerging in many industries today to develop and deploy services that cater to individual needs and preferences. Music afficianados can create their own radio stations from Pandora~\cite{pandora}, bibliophiles can receive 
book recommendations from goodreads.com~\cite{goodreads}, and Google will provide directions between any two points with warnings of delays in real-time, as well as allowing users to choose the mode of transportation~\cite{maps}.
These services leverage large databases to learn and extract information relevant to individuals.  
A class of techniques that transforms data into actionable information goes by the name of Machine Learning (ML)\cite{pythonmachinelearning}.
ML has recently become a popular method to answer questions and solve problems that are too complex to solve via traditional methods.

The primary objective of this study is to show how ML models can be trained using publically available data to produce personalized survival prognosis curves. The methods presented below can be applied to any type of temporal outcome data, including survival, cost, complication and toxicity data. Traditionally, cancer survival curves have been estimated using Kaplan-Meier methods~\cite{cam}. Kaplan-Meier methodology also uses large datasets to make predictions, but the resulting curves are summaries for a population and not necessarily relevant or accurate for any given individual. This property of Kaplan-Meier methods is exacerbated when dealing with heterogeneous populations~\cite{Kumbasar2016140}.  
The capability to provide individualized survival curve prognoses is a direct result of recent advances in computing power and ML algorithms. Similar methodology is becoming commonplace in many industries.
 These techniques are now infiltrating the healthcare industry.

The  Surveillance, Epidemiolgy, and End Results (SEER) Program of the National Cancer Institute (NCI) program is the most recognized authoritative source of information on cancer incidence and survival in the United States and is the primary data source for this study. SEER currently collects and publishes cancer incidence and survival data from population-based cancer registries covering approximately 28 percent of the US population.
The SEER Program  has been collecting data since 1973.
Intuitively 
researchers feel confident
 that this data will surface information crucial to patients and providers, including the relationships between the collected data (demographics, staging, treatment and disease characteristics) and survival outcomes.
Though these relationships evade capture by traditional methods, it is possible to surface them with two machine learning techniques known as random forests and neural networks.

One challenge of the SEER data that is shared by many survival datasets is the inclusion of censored data.
Observations are labeled censored when the survival information is incomplete.
 The SEER data contains the number of months each patient survived, as well as the vital status.
Traditional methods to deal effectively with this kind of "right-censored data'' include Kaplan-Meier curves
and Cox Proportional Hazard models \cite{cam}.

Previous work applying machine learning methods to subsets of the SEER data include creative attempts to deal with the problems presented by right-censored data. Shin et al.~\cite{ISI:000337467400005} use semi-supervised learning techniques to predict 5 year survival, essentially imputing values for SEER records where the survival infomation is censored at a value less than 5 years. Zolbanin et al.~\cite{ISI:000355882700012} remove all records corresponding to patients who were living but censored within the 60 month study window. This treatment biases the predictions and leads to overly pessimistic predictions.

Previous work applying machine learning methods based on decision trees to survival data in general have a long history, starting with Gordon et al.~\cite{Gordon19851065}. A summary of more recent developments concerning survival trees is provided by Bou-Hamad et al.~\cite{Bou-Hamad201144}. These methods focus on altering the splitting critieria used in decision tree growth to account for the censoring, and use Kaplan-Meier methods at the resulting nodes for prediction purposes. These methods do not generalize to non-tree-based machine learning algorithms, though Ishwaran et al. have extended the methodology to random survival forests, ensembles of survival trees~\cite{Ishwaran20101056}.

 Instead of modifying existing learning algorithms, we focus attention on the input data. This approach allows us to take advantage of powerful and rapidly improving machine learning derived discrete classifiers without modification. 
 The essential idea is to recast the problem as a discrete classification problem (predicting the liklihood that a patient is alive in any given month) instead of a regression problem (predicting survival months). Treating months after diagnosis as a discrete feature, the SEER data (or any other right-censored data) can be transformed to make predictions for the hazard function (the probability of dying in the next month, given that the patient has not yet died).
The survival function can then be derived from the hazard function.

\section{Materials and Methods}
\label{sec:materialsandmethods}

\subsection{Data preparation and preprocessing}
\label{subsec:dataprep} 

For this study we use the publically available 1973-2012 SEER incidence data files corresponding to colon, breast and lung cancer. These files are listed in subsection~(\ref{S1_Text}).
A great deal of data munging is necessary before using these SEER incidence files as input into machine learning algorithms. 
The input data was recoded and reshaped to comply with the requirements of the analysis program. Details are included in subsection~(\ref{S2_Text}). Biefly, we transformed the location variables that are given as categorical State and County code pairs
to (latitude, longitude, elevation) triples using the Google Maps API, as well as one-hot encoded all categorical variables.

In the SEER data, there is a record for each primary tumor. If multple records exist for a given patient, only the first chronologically was included. The full set of conditions defining the subsets of the SEER data used in this study is included in section~(\ref{S3_Text}).

Before applying machine learning models trained with these datasets, we describe in detail a method that takes full advantage of all the data, including the right-censored data, and which involves a simple and intuitive transformation, culminating in the full set of features and target variable listed in sections~(\ref{S4_Text},~\ref{S5_Text},~\ref{S6_Text}). 

\subsection{Transformation of Censored Data for Machine Learning}
\label{subsec:transformation}

In this section we describe a transformation of right-censored data. The tramsformed data can be used as input to machine learning algorithms which learn the hazard fuction. The full details of this transformation, and a large inspiration for this study, can be found in this blog post~\cite{kuhn}.

The key observation is to note that the hazard function at any given time point can be directly learned via standard machine learning methods. The hazard function can be written as
\begin{equation}
\label{eq:hhazard}
\lambda(\mathbf{X}_{i}, t_{j}) = P(Y = t_{j}|Y \geq t_{j}, \mathbf{X}_{i}),
\end{equation}
the probability that, if someone has survived up until month $t_{j}$, they will die in that month.
 $j$ runs from 0 to 107, and $\mathbf{X}_{i}$ corresponds to the single row corresponding to patient $i$ in the original untransformed dataset.
107 months was the maximum value of survival months in all three of the cancer datasets, and is a consequence of the data subsets chosen for this study.
$Y$ represents the true, uncensored number of survival months of the patient.
What is actually provided in the SEER data is the related variable \codewhite{SURVIVAL MONTHS} $T$ (how long each subject was in the study), and whether they exited by dying or being censored ($D$), \codewhite{VITAL STATUS RECODE}. 
$D$ is a Boolean variable, so $D = 1$ if $T = Y$, and $D = 0$ if $T < Y$.

It follows directly from equation~(\ref{eq:hhazard}) that 
\begin{equation}
\label{eq:pmf}
P(Y = t_{j} | \mathbf{X}_{i}) = \lambda(\mathbf{X}_{i}, t_{j}) \prod_{k=1}^{j-1} (1 - \lambda(\mathbf{X}_{i}, t_{k}))
\end{equation}
Knowing $P(Y = t_{j} | \mathbf{X}_{i}) $ for all $t_{j}$ gives the 
full probablity distribution of dying at time Y~\cite{kuhn}.
The survival function is then readily derived from this distribution as
\begin{equation}
\label{eq:cdf}
S(\mathbf{X}_{i},t_{k}) = 1 - CDF(\mathbf{X}_{i}, t_{k})
\end{equation}
where $CDF(\mathbf{X}_{i}, t_{k}) = \sum_
{j=1}^{k} P(Y = t_{j} | \mathbf{X}_{i}) $ is the cumulative density function correponding to the probability mass function in equation~(\ref{eq:pmf})~\cite{downey}.

Treating $T$ as just another covariate is the key to the transformation. Each datapoint in the hidden classification problem is the combination of an $\mathbf{X}_{i}$ in the orginal dataset plus some month $t_{j}$, and the classification problem is "did point $\mathbf{X}_{i}$ die in month $t_{j}$.'' We will call this new variable $D_{ij}$ (\codewhite{newtarget}).
We can transform our original dataset into a new one, with one row for each month that each $\mathbf{X}_{i}$ is in the sample; train a standard classifier on this new dataset with $D_{ij}$ as the target, and derive a survival model from the orginal dataset.
Psuedocode for this transformation is found in section~(\ref{S7_Text}).

Explicit examples will help make this transformation clear.
The untransformed records represented in section~(\ref{S1_Table}) are transformed to the multiple records shown in section~(\ref{S2_Table}).

One obvious side effect of this transformation is that it increases the length of the dataset.
For this study, the original, untransformed colon cancer DataFrame has shape $(113072, 103)$, and the total transformed colon cancer DataFrame has shape $(4165251, 103)$.
Similary, the original, untransformed lung cancer DataFrame has shape $(177089, 115)$, and the total transformed lung cancer DataFrame has shape $(3079931, 115)$.
The biggest increase in dataset size occured with the breast cancer data, which is a consequence of the relatively high survival rates in breast cancer. A patient who is censored with a recorded survival months of 48 will contribute 49 rows to the transformed dataset.  
The original, untransformed breast cancer DataFrame has shape $(329949, 67)$, and the total transformed breast cancer DataFrame has shape $(15085711, 67)$.
Training machine learning algorithms on such large datasets, even after splitting into training and testing sets described below, requires large RAM. All computations for this study were performed on a Dell XPS 8700 Desktop with 32GB of RAM. The training times involved in the classification task of learning the hazard function $\lambda(\mathbf{X}_{i}, t_{j})$ for the chosen model parameters were on the order of a few hours or less, but the evaluation of the AUC performance metrics associated with the 6, 12, and 60 month binary survival classifiers
 took more than 24 hours for the random forest models. These AUC performance metrics provided the feedback mechanism to adjust the model hyperparameters.
\subsection{Training and Test Partitions}
\label{subsec:traintest}

The datasets were split into training and test sets at the patient level, with 97\% of patients assigned to the training set, and the remaining 3\% of patients assigned to the test set.
All records corresponding to a given patient were assigned exclusively to either the training or test set.
This choice of an unusually low percentage of data in the test set was made for two reasons. The performance metrics described in section~(\ref{sec:performancemetrics}) for the given choice of training and test partition of the data took well over 24 hours for the random forest models; choosing the conventional 80/20 split would have resulted in prohibitively long times for the traning-performance metric feedback loop. Because of the large size of the data set, this choice of training and test partition still leads to an acceptably large test set for the purposes of model evaluation.
An additional characteristic of this transformed data that requires careful treatment involves balancing. The transformation results in many new records with the target variable \codewhite{newtarget} == 0. The training and test sets must be chosen such that the ratio of the number of records with \codewhite{newtarget} == 0 to that of the number of records with \codewhite{newtarget} == 1 is the same in the training and test datasets.
This ratio turns out to be $\approx 396$ for the breast cancer data, $\approx  99$ for the colon cancer data, and 
$\approx 22.75$ for the lung cancer data. 
The shapes of the training and testing datasets for breast cancer used in this study are $(14936862, 67)$ and 
$(148849, 67)$, respectively.
For lung cancer, the corresponding datasets have shapes $(2988768, 115)$ and $(91163, 115)$.
Finallly, for colon cancer the partition into training and test datasets of the transformed data have the shapes 
$(3958008, 103)$ and $(207243, 103)$. Multiple rows correspond to the same test patient in these datasets.
The colon cancer test dataset represents 5654 distinct patients; the breast cancer test dataset represents 3300 distinct patients; and the lung test dataset contains data for 5313 distinct patients.

The models described below are trained to learn the values of \codewhite{newtarget}, which is a binary variable: a value of 0 indicates that the subject is still alive at the given month, while a value of 1 indicates that the patient died at that particular value of \codewhite{months}. The random forests and neural networks described below are binary classifiers with the target \codewhite{newtarget}. 
Both the random forests and neural networks are capable of not only performing strict class prediction, i.e. predicting whether \codewhite{newtarget} is 0 or 1, but are also able to predict the probability of \codewhite{newtarget} being 0 or 1, and are thus able to learn the hazard function.

\subsection{Prediction Models}
\label{sec:predmodels}

With the datasets transformed as described above, we are now able to use them to train and evaluate machine learning classifiers.
The classifier models described in this section are learning the hazard function: given all of the data given in sections~(\ref{S4_Text},~\ref{S5_Text},~\ref{S6_Text}) for each cancer type, which includes the field \codewhite{months} (the months after diagnosis), the models predict the target variable \codewhite{newtarget}, which is a binary class label equal to 1 if the subject died in that month and 0 otherwise.

From the hazard function for each unique patient, we can construct the
survival function as in Equation~(\ref{eq:cdf}).
The relevant python code is available at the github repository containing supplemental material for this study~\cite{supp}.
For each patient $i$, all input data minus \codewhite{months} and \codewhite{newtarget} is represented by $\mathbf{X}_{i}$. After the classfier models have trained with target \codewhite{newtarget} on the training set, each subject's survival function is computed in the corresponding test set.
These functions are computed by using the model to predict $\lambda(\mathbf{X}_i, t_{j})$ for $j$ running from 0 to 107 months, and $\mathbf{X}_{i}$ corresponds to the single row corresponding to subject $i$ in the original untransformed dataset.
107 months was the maximum value of survival months in all three of the cancer datasets, and is a consequence of the data subsets chosen for this study.

\paragraph{Decision Trees and Random Forests}
Decision tree classifiers are attractive models because they can be intrepeted easily. Like the name decision tree suggests, we can think of this model as breaking down our data by making decisions based on asking a series of questions.
Based on the features in our training set, the decision tree model learns a series of questions to infer the class labels of the samples. 

Random forests have gained huge popularity in applications of machine learning during the last decade due to their good classification performance, scalability, and ease of use. Intuitively, a random forest can be considered as an ensemble of decision trees. The idea behind ensemble learning is to combine weak learners to build a more robust model, a strong learner, that has a better generalization error and is less susceptible to overfitting. 

The goal behind ensemble methods is to combine different classifiers into a meta-classifier that has a better generalization performance than each individual classifier alone. For example, assuming that we collected predictions from 10 experts, ensemble methods would allow us to strategically combine these predictions by the 10 experts to come up with a prediction that is more accurate and robust than the predictions by each individual expert. The individual decision trees that make an ensemble are called base learners, and as long as the error rate of each base learner is less than .50, the combined random forest will benefit from the affects of combining predictions to achieve a far greater accuracy.

A big advantage of random forests is that honing in on suitable hyperparameter values (the number of trees in the forest, the depth of each decision tree, the specific measure of information gain used to choose the node splitting, etc) is not very difficult. The ensemble method is robust to noise from the individual decision trees, which helps to prevent overfitting (memorizing the training dataset targets instead of generalizing from learned rules to perform successfuly on unseen data). The only parameter that has a clearly noticeable effect on performance is the number of trees to include in the forest; in general, the more trees the better the performance, but there is a price to pay in terms of computational cost. The number of trees for the forests trained in this study was relatively small, 20 trees for breast cancer and 25 for both the lung and colon cancer models. We have used the Python scikit-learn implemenation of the Random Forest machine 
learning classifier~\cite{rf}.
Random Forests are frequent winners of the Kaggle machine learning competitions~\cite{kagglerf}.
The model parameters for each cancer type are given in
sections~(\ref{S8_Text},~\ref{S9_Text},~\ref{S10_Text}).


\paragraph{Multi-Layer Perceptron Neural Networks}
Neural networks are a biologically-inspired programming paradigm that enables computers to learn from observational data~\cite{deeplearning}.
The pharmaceutical industry recently started to use deep learning techniques for drug discovery and toxicity prediction, and research has shown that these novel techniques substantially exceed the performance of traditional methods for virtual screening~\cite{toxicity}.

We have used the Multi-Layer Perceptron Neural Network (MLP neural network) implementation Keras developed at MIT.
Keras was initially developed as part of the research effort of project ONEIROS (Open-ended Neuro-Electronic Intelligent Robot Operating System)~\cite{keras}.
Keras is a minimalist, highly modular neural networks library, written in Python and capable of running on top of either TensorFlow or Theano. The model architecture for each cancer type is given in sections~(\ref{S11_Text},~\ref{S12_Text},~\ref{S13_Text}).

\section{Results}
\label{sec:results}

\begin{figure}[!ht]
\centering 
\includegraphics[width=.90\textwidth,origin=c]{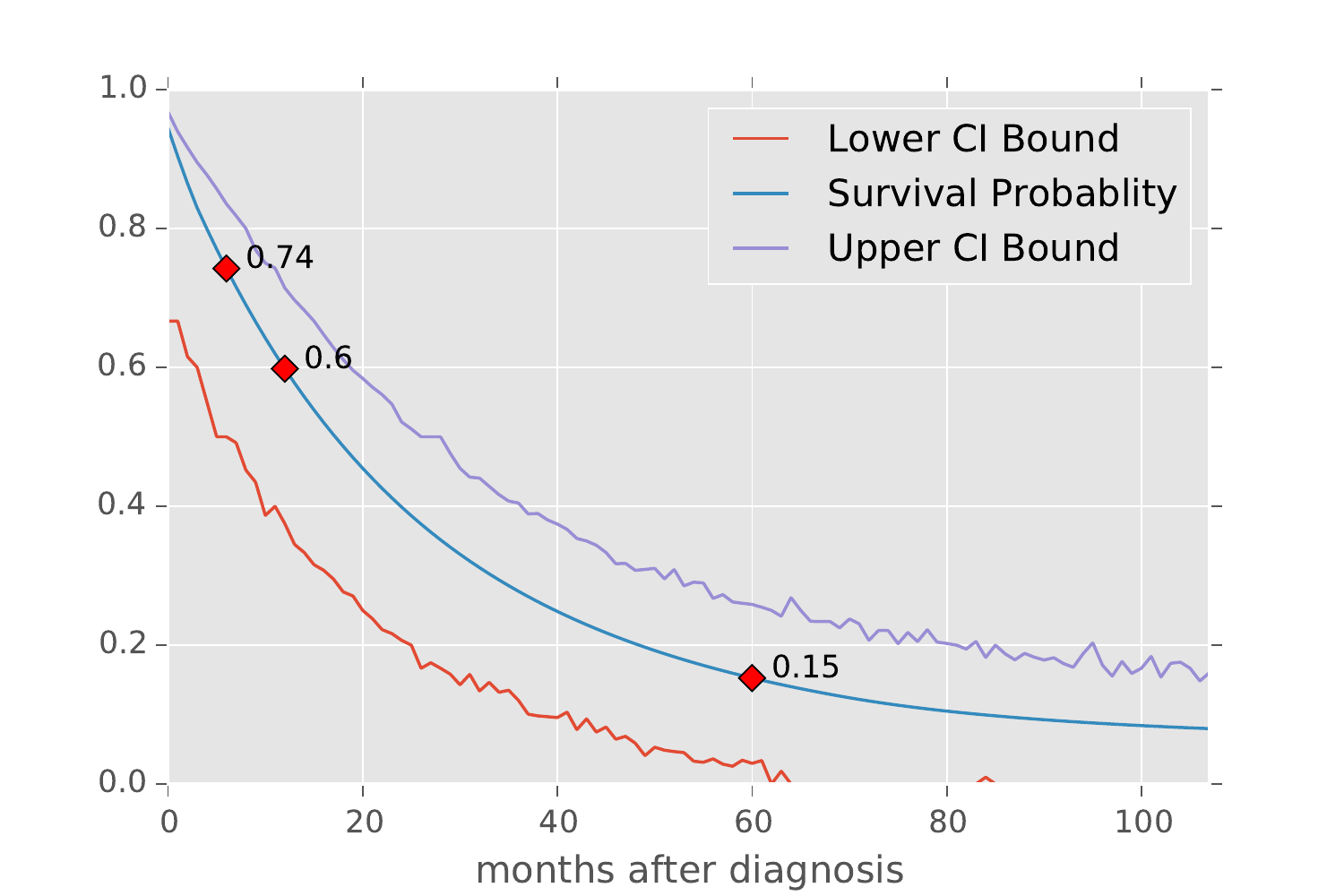}
\caption{\label{fig:survivalexample} Example of the construction of the binary classifiers for 6, 12, and 60 months survival. A patient's hazard curve $\lambda(\mathbf{X}_{i}, t_{j})$ is predicted by the model for times out to 107 months. The survival curve is then readily computed as in Equation~(\ref{eq:cdf}). For this example, the 6-month and 12-month classifiers predict survival, while the 60-month classifier predicts death.}
\end{figure}

In order to evaluate the performance of the models, we first construct three binary classifiers corresponding to whether or not a subject survived 6, 12, or 60 months after diagnosis. 
We iterate over all distinct patient indices in the test set, compute the predicted survival function, and capture the values corresonding to 6, 12, and 60 months.  
If the survival function evaluted at 6 months is greater than or equal to .5 for a given patient, then the 6 months binary classifier predicts that that patient will be alive 6 months after diagnosis. Similarly, if the survival function evaluted at 12 months is less than .5, then the 12 months binary classifier predicts that that subject will be dead 12 months after diagnosis. Fig.~(\ref{fig:survivalexample}) illustrates the method; in this case the 6-month and 12-month classifiers predict survival, while the 60-month classifier predicts death.

Because of censoring it is necessary to apply some Boolean filters to the data in order to correctly assess the resulting classifiers.
To construct AUC curves for the 6 month classifier, we restrict ourselves to 
patients in the test data where either of the following mutually exlusive conditions holds:

\begin{itemize}[noitemsep]
\item \codewhite{survival\_months} $>=$ 6 AND \codewhite{vital\_status\_recode} == 0
\item \codewhite{vital\_status\_recode} == 1
\end{itemize}

That is, we restrict ourselves to subsets of the data where we know for certain whether or not the subject survived at least 6 months. Similarly for the 12 and 60 months surivival classifiers. 

\paragraph{Survival Curve Error Estimates}
The following bootstrap method was used to calculate the upper and lower bounds corresponding to 95\% confidence intervals.
From equation~(\ref{eq:cdf}), we can obtain the cumulative distribution function (CDF) associated with each individual survival curve.
We then sample from this CDF in a way that reflects the underlying data used to produce the model. The training data used to create the model has an underlying distribution of survival months. In the transformed training dataset, each subject contributes as many rows as the number of survival months plus one (patients with zero survival months still contribute one row to the training data). A patient that survived 50 months contributes 51 ``points" to the training of the model. If all patients lived out to 107 months, the model would contain less uncertainty. This observation leads to the following algorithm for determining the error estimates to the predicted survival curves:

\begin{itemize}[noitemsep]
\item compute the CDF associated with the survival curve
\item use the underlying training data CDF of survival months to choose the number of points
to draw from the survival curve CDF, and compute a new survival curve 
\item Repeat the previous step 10,000 times and collect the curves into a list. Changing the number of curves affects how smooth the upper and lower bounds are, but does not affect the interval size between for each month.
\item extract for each month from the list of curves the .975 and .025 percentiles to record the values for the upper and lower curves
\end{itemize}

The process is analogous to the following hypothetical situation. Imagine a patient going to an expert to get a survival prognosis.
After collecting data on the patient and keeping records, the expert predicts the central, single survival curve. The patient then seeks multiple ``second opinions.'' These second opinions are generated not from independent examinations of the patient, but by outside experts sampling from the data already collected by the first expert.
Then the predictions of 95\% of these 10,000 experts all fall within the band determined by the upper and lower curves.

\subsection{Performance Metrics}
\label{sec:performancemetrics}

\paragraph{AUC scores}
The AUC scores for each of the 18 different binary classifiers are listed in section~(\ref{S3_Table}). 
The lowest AUC in section~(\ref{S3_Table}) is .765, corresponding to the lung neural network model predictions for 6 months survival, while the highest AUC in section~(\ref{S3_Table}) is .885, corresponding to the breast random forest model predictions for 12 months survival.


\paragraph{Model Agreement}
An additional means of validating the predictions of these models is by comparing their predictions to each other for the same set of input data. 
section~(\ref{S4_Table}) shows the strong agreement between the random forest and neural network classifiers for each cancer type. Python code showing how the values in section~(\ref{S4_Table}) are computed is available in the files 
\codewhite{NewPatientBreastCF.html}, \codewhite{NewPatientColonCF.html}, and \codewhite{NewPatientLung.html} in the GitHub repository containing supplemental matierial for this study~\cite{supp}. Section~(\ref{S4_Table}) is computed as follows. 
For each cancer type (breast,colon, and lung), do the following:

\begin{itemize}[noitemsep]
\item use the corresponding Random Forest and Neural Network models to compute the survival curves for all of the test subjects
\item extract the values of the survival curve evaluted for 6, 12, and 60 months for both models
\item if both models predict less than .5 or both models predict greater than or equal to .5, that counts as agreement
\item otherwise, the models disagree
\end{itemize}

\begin{figure}[tbp]
\begin{adjustwidth}{-2.25in}{0in} 
\centering
\includegraphics[width=1.2\textwidth,origin=c]{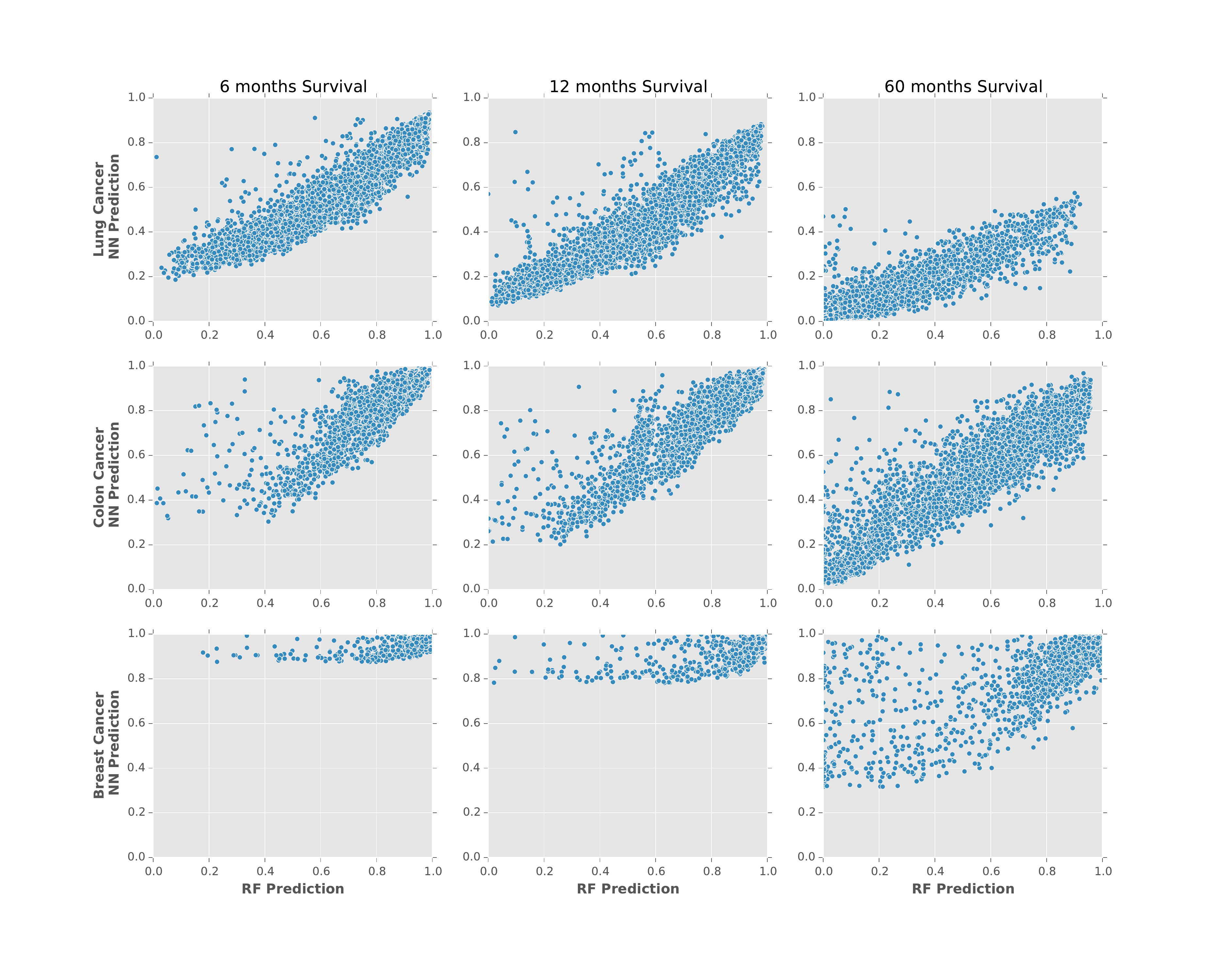}
\caption{\label{fig:totalscatter} Scatter plots showing the correlations between the MLP model's prediction and RF model's prediction for the probablity of surviving at least 6, 12, and 60 months for the lung, colon and breast cancer test data.}
\end{adjustwidth}
\end{figure}

Fig.~(\ref{fig:totalscatter}) shows scatter plots of the neural network and random forest predictions for 6, 12, and 60 month survival prediction for each cancer type. The correlations of the neural network and random forest predictions are listed in section~(\ref{S5_Table}). Event though breast cancer 6 month survival has the highest percentage agreement between the two classifiers (over 99$\%$), it also has the lowest correlation (.676). This seeming contradiction is a result of the high 6-month survival rate for breast cancer, and that the random forest and neural network models seem to treat this case differently. The neural network models in general appear to be less pessimistic in general than the random forest model predictions.

The high level of agreement between two models lends confidence to the notion that they have both learned from the training data and are generalizing well. 

\subsection{Survival Curve Prediction Apps}
\label{sec:apps}

The six models have their full hyperparameter and architecture presented in sections~(\ref{S8_Text},~\ref{S9_Text},~\ref{S10_Text},~\ref{S11_Text},~\ref{S12_Text},~\ref{S13_Text}). Python code for all six model training and evaluation is available at the githib respository containing supplemental material for this study~\cite{supp}.

Using the popular Flask microframework for web applications~\cite{flask}, we have made web applications corresponding to the six models. The list of web applications below will allow readers to freely experiment with the models.

\begin{enumerate}[noitemsep]
\item breast cancer 
    \begin{enumerate}[noitemsep]
    \item random forest: \url{https://github.com/doolingdavid/breast-cancer-rf-errors.git}
    \item neural network: \url{https://github.com/doolingdavid/breast-cancer-nn-errors.git}
    \end{enumerate}
\item lung cancer
   \begin{enumerate}[noitemsep]
   \item random forest: \url{https://github.com/doolingdavid/lung-cancer-rf-errors.git}
   \item neural network: \url{https://github.com/doolingdavid/lung-cancer-nn-errors.git}
    \end{enumerate}
\item colon cancer
  \begin{enumerate}[noitemsep]
   \item random forest: \url{https://github.com/doolingdavid/colon-cancer-rf-errors.git}
   \item neural network: \url{https://github.com/doolingdavid/colon-cancer-nn-errors.git}
   \end{enumerate}
\end{enumerate}

After downloading the .zip file associate with one of the above web applications, and assuming 
python is installed on your system, you can launch the application by running
\begin{verbatim}
>python hello.py
\end{verbatim}
and pointing the browser to the local server: \codewhite{http://127.0.0.1:5000}, or 
\codewhite{http://localhost:5000}.

\begin{figure}[!ht]
  \centering
    \includegraphics[scale=.8]{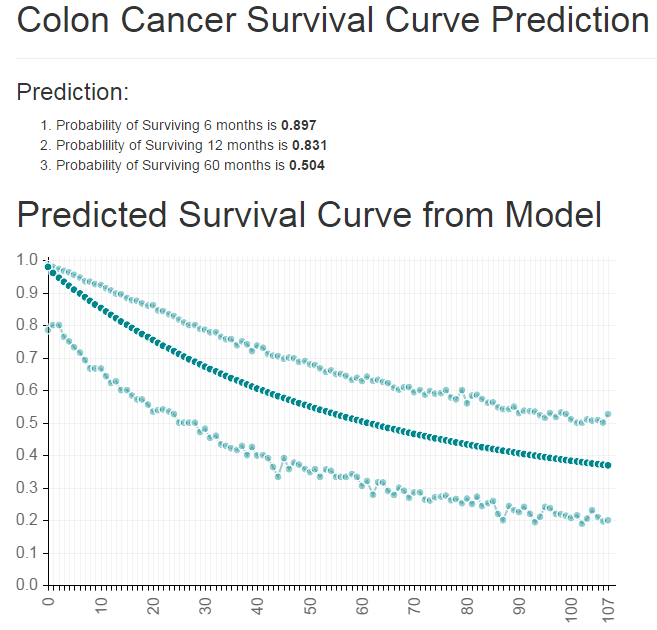}
\caption{\label{fig:boston1940} Colon Cancer Survival Curve predicted from the data in 
section~(\ref{S6_Table}) using the neural network web app \url{https://github.com/doolingdavid/colon-cancer-nn-errors.git}.}
\end{figure}

 For example, using the Colon Cancer neural network app, and 
inputing the values listed in section~(\ref{S6_Table}) results in the survival curve depicted in Figure~(\ref{fig:boston1940}); the predicted probablities of living 
at least 6, 12, and 60 months are .89, .83, and .50, respectively.


Changing the data in section~(\ref{S6_Table}) so that the address field is changed from Boston, Massachusetts to Denver, Colorado but keeping all other variables unchanged results in the predicted probabilities of living at least 6, 12, and 60 months: .945, .902, .665. 
Behind the scenes, the apps use the input to the address field to make a call to the Google Maps API to convert the address into a latitude, longitude and elevation.
These probablities are noticeably higher and reflect the documented effects of both longitude and elevation on cancer treatment and prognosis in the United States~\cite{kob4}.

A similar example of how changing the inputs to the models affects the predicted survival curves in interesting ways can be seen with the random forest model for lung cancer. Changing the data in section~(\ref{S7_Table} by toggling between the male/female, and married/single four possible permutations results in the following prediction probabilites for 6, 12, and 60 month survival:

\begin{itemize}[noitemsep]
\item male/married:  .53, .27, .01
\item male/single: .35, .18, .009
\item female/married:  .55, .31, .01
\item female/single: .50, .27, .01
\end{itemize}


Inputting the same combinations of data into the lung cancer neural network app \url{https://github.com/doolingdavid/lung-cancer-nn-errors.git}
yields the following probabilities:

\begin{itemize}[noitemsep]
\item male/married: .42, .24, .04
\item male/single: .40, .22, .03
\item female/married: .44, .26, .04
\item female/single: .42, .24, .04
\end{itemize}

It it interesting to note that both the random forest and neural network lung cancer models predict greater 6 month survival rates for married people, with a slightly greater benefit for males than females. The effect is greater in the random forest model, but is also visible in the neural network model.

\section{Discussion}
\label{sec:discussion}

The purpose of this study has been twofold; to develop a general methodology of data transformation to survival data with censored observations so that machine learning algorithms can be applied and to apply the methodology to create models of personalized survival curve prognosis.
To help further refine the methodology, we would like to apply it to different survival datasets~\cite{umass}, not necessarily within the healthcare domain. In particular, the methods presented in this paper do not take into account time varying features. For example, the \codewhite{cs\_tumor\_size} variable that has been a part of this study is kept fixed at the value measured at diagnosis for all records corresponding to a given subject. Clearly, the actual tumor size varies along with time and a sophisitcated model can be developed to take this into account, given available datasets.

The SEER database has been linked with claims data in the SEER-Medicare Linked Database~\cite{seermed}. This linkage allows for the identification of additional clinical data for each record in the SEER database and allows for an enrichment of the models presented in this study, and is an avenue for further investigation.

An additional avenue of research concerns the broad concept of causality. As demonstrated in section~(\ref{sec:apps}), there appears to be a correlation between marital status and survival prognosis. Does this mean that if a single person in Boston, Massachusetts is diagnosed with cancer, that they should immediately get married and move to Denver? Of course not. 
But personal discussions with providers has confirmed for one of the authors (D.D.) that married males tend to be much more diligent in following instructions than their single counterparts. What appears to be in effect is that some of the SEER data is providing an identifiable signature of underlying causes not directly represented by the data. Latent variables not directly seen in the data are still providing echos of patterns in the data and the sheer volume allows us to see glimpses of these patterns. Marital status is in some instances a surrogate for the presence of a strong social structure and support group surrounding a patient, which presence presumably leads to more desirable survival prognosis. 
The daunting and exciting task of teasing out actual causality relationships within machine learning contexts has been pioneeered by Judea Pearl of the University of California, Los Angeles and seems particulary relevant and applicable to censored survival data. Combining the methodology presented in this study with that of the pioneering work of Judea Pearl on causality will be a fruitful avenue for future research.

\section{Supporting Information}
\label{sec:supporting}

\subsection{S1 Text}
\label{S1_Text}

Raw SEER datafiles

\begin{itemize}[noitemsep]
\item incidence\textbackslash yr1973\_2012.seer9\textbackslash COLRECT.txt
\item incidence\textbackslash yr1973\_2012.seer9\textbackslash BREAST.txt
\item incidence\textbackslash yr1973\_2012.seer9\textbackslash RESPIR.txt
\item incidence\textbackslash yr1992\_2012.sj\_la\_rg\_ak\textbackslash COLRECT.txt
\item incidence\textbackslash yr1992\_2012.sj\_la\_rg\_ak\textbackslash BREAST.txt
\item incidence\textbackslash yr1992\_2012.sj\_la\_rg\_ak\textbackslash RESPIR.txt
\item incidence\textbackslash yr2000\_2012.ca\_ky\_lo\_nj\_ga\textbackslash COLRECT.txt
\item incidence\textbackslash yr2000\_2012.ca\_ky\_lo\_nj\_ga\textbackslash BREAST.txt
\item incidence\textbackslash yr2000\_2012.ca\_ky\_lo\_nj\_ga\textbackslash RESPIR.txt
\item incidence\textbackslash yr2005.lo\_2nd\_half\textbackslash COLRECT.txt
\item incidence\textbackslash yr2005.lo\_2nd\_half\textbackslash BREAST.txt
\item incidence\textbackslash yr2005.lo\_2nd\_half\textbackslash RESPIR.txt
\end{itemize}

\subsection{S2 Text}
\label{S2_Text}

A preprocessing step common to each of the three cancer types studied involves the SEER \codewhite{STATE-COUNTY RECODE} variable.
The \codewhite{STATE-COUNTY RECODE} field is a state-county combination where the first two characters represent the state FIPS code and the last three digits represent the FIPS county code.  The FIPS code is a five-digit Federal Information Processing Standard (FIPS) code which uniquely identifies counties and county equivalents in the United States, certain U.S. possessions, and certain freely associated states.
This particular field illustrates an important characteristic of machine learning, that is, the difference  between \textit{categorical features} and \textit{numeric features}. All input into a machine learning algorithm must be numeric, but real numbers carry with them the usually extremely useful property known as the well-ordering property. Machine learning algorithms use the well-ordering property of the real numbers to learn.
But if one is tasked with encoding a categorical feature into suitable numeric format for machine learning, it is necessary to do so in a way that removes the well-ordering property. Categorical variables are commonly encoded using one-hot encoding, in which the explanatory variable is encoded using one binary feature for each of the variable's possible values~\cite{bowles}.

One-hot encoding needs to be applied to all of the nominal categorical variables in the SEER data that we wish to include in our predictive models.
In particular, in order to include the geophgraphical information contained in the SEER categorical variable \codewhite{STATE-COUNTY RECODE}, it becomes necessary to create a new feature variable for each of the distinct (state,county) pairs in the data. In the United States, there are approximately 3,000 counties. Clearly, transforming the \codewhite{STATE-COUNTY RECODE} data representation into distinct (state$\_$county) columns will explode the dataset to become wider than is optimal for machine learning. Adding extra columns to your dataset, making it wider, requires more data rows (making it taller) in order for machine learning algorithms to effectively learn~\cite{bowles}. Because one-hot coding \codewhite{STATE-COUNTY RECODE} would cause such drastic shape changes in our data, we wish to avoid doing so. Fortunately, this variable, though given as a categorical variable, is actually a recode for three ordinal variables. There is an ordering among the (state$\_$county) columns, namely longitude, latitude, and elevation. We can transform the data in \codewhite{STATE-COUNTY RECODE} into three new numerical columns: \codewhite{lat}, \codewhite{lng}, and \codewhite{elevation}.

For example, section~(\ref{S8_Table}) shows how five entries of \codewhite{STATE-COUNTY RECODE} corresponding to counties within New Mexico can be represented by the 
\codewhite{elevation}, \codewhite{lat}, and \codewhite{lng} features.


It is a simple exercise to construct the full lookup table from the SEER \\  \codewhite{STATE-COUNTY RECODE} variable to the corresponding three values \codewhite{elevation}, \codewhite{lat}, and \codewhite{lng}. We use the publically available datafile from the United States Census Bureau~\cite{census} to map the state FIPS and county FIPS codes to query strings like those in the \codewhite{address} field in section~(\ref{S8_Table}). 
It is then possible to programmatically query the Google Maps Geocoding API for the latitude and longitude~\cite{geocode}, and the Google Maps Elevation API for the corresponding elevation~\cite{elevation}.
An added benefit of this shift from the single categorical variable \codewhite{STATE-COUNTY RECODE} to the three continuous numerical variables \codewhite{lat}, \codewhite{lng}, and \codewhite{elevation} is that input into the web applications described later are not restricted to the states and counties coverered in the SEER registries; in fact, the input to the models can be any address you would enter into Google Maps and calls to the Google Maps Geocoding API and the Google Maps Elevation API provide the conversion from the address string to the input variables \codewhite{lat}, \codewhite{lng}, and \codewhite{elevation}. The full lookup table analogous to section~(\ref{S8_Table}) is available from a GitHub repository containing supplemental information for this study~\cite{supp}.

\subsection{S3 Text}
\label{S3_Text}

 The four COLRECT.txt files were imported into a pandas DataFrame object.
This data was then filtered according to the conditions in section~(\ref{S9_Table}).
The RESPIR.txt and BREAST.txt files were imported into separate dataframes in similar fashion and filtered according
to the conditions in section~(\ref{S10_Table}) and section~(\ref{S11_Table}), respectively.
The SEER variable \codewhite{CS TUMOR SIZE} records the tumor size in millimeters if known. But if not known, \codewhite{CS TUMOR SIZE} is given as '999', to indicate that the tumor size is "Unknown; size not stated; not stated in pateint record.'' In this study, we discard those records, as indicated in sections~(\ref{S11_Table},~\ref{S9_Table},~\ref{S10_Table}).

The following categorical features were one-hot encoded for each of the three datasets:

\begin{itemize}[noitemsep]
\item \codewhite{SEX},
\item  \codewhite{MARITAL STATUS AT DX},
\item \codewhite{RACE/ETHNICITY},
\item \codewhite{SPANISH/HISPANIC ORIGIN},
\item \codewhite{GRADE},
\item \codewhite{PRIMARY SITE},
\item \codewhite{LATERALITY},
\item \codewhite{SEER HISTORIC STAGE A},
\item \codewhite{HISTOLOGY RECODE--BROAD GROUPINGS},
\item \codewhite{MONTH OF DIAGNOSIS},
\item  \codewhite{VITAL STATUS RECODE},
\end{itemize}
and the \codewhite{STATE-COUNTY RECODE} variable was dropped and replaced with the \codewhite{elevation}, \codewhite{lat}, and \codewhite{lng} variables for all three datasets as illustrated in Table~(\ref{tab:nmhead}).

\subsection{S4 Text}
\label{S4_Text}

{\bf Colon Cancer Feature Selection} The feature set used as input into both the Random Forest and Neural Network models, after the transformation described in section~(\ref{subsec:transformation}) is given below and also available in full detail in the file 
\codewhite{NewPatientColonML.html}.

\begin{itemize}[noitemsep]
\item cs\_tumor\_size
\item elevation
\item grade\_cell type not determined
\item grade\_moderately differentiated
\item grade\_poorly differentiated
\item grade\_undifferentiated; anaplastic
\item grade\_well differentiated
\item histology\_recode\_broad\_groupings\_acinar cell neoplasms
\item histology\_recode\_broad\_groupings\_adenomas and adenocarcinomas
\item histology\_recode\_broad\_groupings\_blood vessel tumors
\item histology\_recode\_broad\_groupings\_complex epithelial neoplasms
\item histology\_recode\_broad\_groupings\_complex mixed and stromal neoplasms
\item histology\_recode\_broad\_groupings\_cystic, mucinous and serous neoplasms
\item histology\_recode\_broad\_groupings\_ductal and lobular neoplasms
\item histology\_recode\_broad\_groupings\_epithelial neoplasms, NOS
\item histology\_recode\_broad\_groupings\_fibromatuos neoplasms
\item histology\_recode\_broad\_groupings\_germ cell neoplasms
\item histology\_recode\_broad\_groupings\_lipomatous neplasms
\item histology\_recode\_broad\_groupings\_miscellaneous bone tumors
\item histology\_recode\_broad\_groupings\_myomatous neoplasms
\item histology\_recode\_broad\_groupings\_neuroepitheliomatous neoplasms
\item histology\_recode\_broad\_groupings\_nevi and melanomas
\item histology\_recode\_broad\_groupings\_paragangliomas and glumus tumors
\item histology\_recode\_broad\_groupings\_soft tissue tumors and sarcomas, NOS
\item histology\_recode\_broad\_groupings\_squamous cell neoplasms
\item histology\_recode\_broad\_groupings\_synovial-like neoplasms
\item histology\_recode\_broad\_groupings\_transistional cell papillomas and carcinomas
\item histology\_recode\_broad\_groupings\_unspecified neoplasms
\item lat
\item laterality\_Left: origin of primary
\item laterality\_Not a paired site
\item laterality\_Only one side involved, right or left origin unspecified
\item laterality\_Paired site, but no information concerning laterality; midline tumor
\item laterality\_Right: origin of primary
\item lng
\item marital\_status\_at\_dx\_Divorced
\item marital\_status\_at\_dx\_Married (including common law)
\item marital\_status\_at\_dx\_Separated
\item marital\_status\_at\_dx\_Single (never married)
\item marital\_status\_at\_dx\_Unknown
\item marital\_status\_at\_dx\_Unmarried or domestic partner
\item marital\_status\_at\_dx\_Widowed
\item month\_of\_diagnosis\_Apr
\item month\_of\_diagnosis\_Aug
\item month\_of\_diagnosis\_Dec
\item month\_of\_diagnosis\_Feb
\item month\_of\_diagnosis\_Jan
\item month\_of\_diagnosis\_Jul
\item month\_of\_diagnosis\_Jun
\item month\_of\_diagnosis\_Mar
\item month\_of\_diagnosis\_May
\item month\_of\_diagnosis\_Nov
\item month\_of\_diagnosis\_Oct
\item month\_of\_diagnosis\_Sep
\item number\_of\_primaries
\item race\_ethnicity\_Amerian Indian, Aleutian, Alaskan Native or Eskimo
\item race\_ethnicity\_Asian Indian
\item race\_ethnicity\_Asian Indian or Pakistani
\item race\_ethnicity\_Black
\item race\_ethnicity\_Chinese
\item race\_ethnicity\_Fiji Islander
\item race\_ethnicity\_Filipino
\item race\_ethnicity\_Guamanian
\item race\_ethnicity\_Hawaiian
\item race\_ethnicity\_Hmong
\item race\_ethnicity\_Japanese
\item race\_ethnicity\_Kampuchean
\item race\_ethnicity\_Korean
\item race\_ethnicity\_Laotian
\item race\_ethnicity\_Melanesian
\item race\_ethnicity\_Micronesian
\item race\_ethnicity\_New Guinean
\item race\_ethnicity\_Other
\item race\_ethnicity\_Other Asian
\item race\_ethnicity\_Pacific Islander
\item race\_ethnicity\_Pakistani
\item race\_ethnicity\_Polynesian
\item race\_ethnicity\_Samoan
\item race\_ethnicity\_Thai
\item race\_ethnicity\_Tongan
\item race\_ethnicity\_Unknown
\item race\_ethnicity\_Vietnamese
\item race\_ethnicity\_White
\item seer\_historic\_stage\_a\_Distant
\item seer\_historic\_stage\_a\_In situ
\item seer\_historic\_stage\_a\_Localized
\item seer\_historic\_stage\_a\_Regional
\item seer\_historic\_stage\_a\_Unstaged
\item sex\_Female
\item spanish\_hispanic\_origin\_Cuban
\item spanish\_hispanic\_origin\_Dominican Republic
\item spanish\_hispanic\_origin\_Mexican
\item spanish\_hispanic\_origin\_Non-Spanish/Non-hispanic
\item spanish\_hispanic\_origin\_Other specified Spanish/Hispanic origin (excludes Dominican Repuclic)
\item spanish\_hispanic\_origin\_Puerto Rican
\item spanish\_hispanic\_origin\_South or Central American (except Brazil)
\item spanish\_hispanic\_origin\_Spanish surname only
\item spanish\_hispanic\_origin\_Spanish, NOS; Hispanic, NOS; Latino, NOS
\item spanish\_hispanic\_origin\_Uknown whether Spanish/Hispanic or not
\item year\_of\_birth
\item year\_of\_diagnosis
\item month
\end{itemize}

and 
\codewhite{newtarget} is the target variable, indicating whether or not the subject died in month given by the value of the \codewhite{month} variable.

\subsection{S5 Text}
\label{S5_Text}

{\bf Lung Cancer Feature Selection} The feature set used as input into both the Random Forest and Neural Network models, after the transformation described in section~(\ref{subsec:transformation}) is given below and also available in full detail in the file 
\codewhite{NewPatientLungML.html}.

\begin{itemize}[noitemsep]
\item cs\_tumor\_size
\item elevation
\item grade\_cell type not determined
\item grade\_moderately differentiated
\item grade\_poorly differentiated
\item grade\_undifferentiated; anaplastic
\item grade\_well differentiated
\item histology\_recode\_broad\_groupings\_acinar cell neoplasms
\item histology\_recode\_broad\_groupings\_adenomas and adenocarcinomas
\item histology\_recode\_broad\_groupings\_blood vessel tumors
\item histology\_recode\_broad\_groupings\_complex epithelial neoplasms
\item histology\_recode\_broad\_groupings\_complex mixed and stromal neoplasms
\item histology\_recode\_broad\_groupings\_cystic, mucinous and serous neoplasms
\item histology\_recode\_broad\_groupings\_ductal and lobular neoplasms
\item histology\_recode\_broad\_groupings\_epithelial neoplasms, NOS
\item histology\_recode\_broad\_groupings\_fibroepithelial neoplasms
\item histology\_recode\_broad\_groupings\_fibromatuos neoplasms
\item histology\_recode\_broad\_groupings\_germ cell neoplasms
\item histology\_recode\_broad\_groupings\_gliomas
\item histology\_recode\_broad\_groupings\_granular cell tumors \& alveolar soft part sarcomas
\item histology\_recode\_broad\_groupings\_lipomatous neplasms
\item histology\_recode\_broad\_groupings\_miscellaneous bone tumors
\item histology\_recode\_broad\_groupings\_miscellaneous tumors
\item histology\_recode\_broad\_groupings\_mucoepidermoid neoplasms
\item histology\_recode\_broad\_groupings\_myomatous neoplasms
\item histology\_recode\_broad\_groupings\_myxomatous neoplasms
\item histology\_recode\_broad\_groupings\_nerve sheath tumors
\item histology\_recode\_broad\_groupings\_neuroepitheliomatous neoplasms
\item histology\_recode\_broad\_groupings\_nevi and melanomas
\item histology\_recode\_broad\_groupings\_osseous and chondromatous neoplasms
\item histology\_recode\_broad\_groupings\_paragangliomas and glumus tumors
\item histology\_recode\_broad\_groupings\_soft tissue tumors and sarcomas, NOS
\item histology\_recode\_broad\_groupings\_squamous cell neoplasms
\item histology\_recode\_broad\_groupings\_synovial-like neoplasms
\item histology\_recode\_broad\_groupings\_thymic epithelial neoplasms
\item histology\_recode\_broad\_groupings\_transistional cell papillomas and carcinomas
\item histology\_recode\_broad\_groupings\_trophoblastic neoplasms
\item histology\_recode\_broad\_groupings\_unspecified neoplasms
\item lat
\item laterality\_Bilateral involvement, lateral origin unknown; stated to be single primary
\item laterality\_Left: origin of primary
\item laterality\_Not a paired site
\item laterality\_Only one side involved, right or left origin unspecified
\item laterality\_Paired site, but no information concerning laterality; midline tumor
\item laterality\_Right: origin of primary
\item lng
\item marital\_status\_at\_dx\_Divorced
\item marital\_status\_at\_dx\_Married (including common law)
\item marital\_status\_at\_dx\_Separated
\item marital\_status\_at\_dx\_Single (never married)
\item marital\_status\_at\_dx\_Unknown
\item marital\_status\_at\_dx\_Unmarried or domestic partner
\item marital\_status\_at\_dx\_Widowed
\item month\_of\_diagnosis\_Apr
\item month\_of\_diagnosis\_Aug
\item month\_of\_diagnosis\_Dec
\item month\_of\_diagnosis\_Feb
\item month\_of\_diagnosis\_Jan
\item month\_of\_diagnosis\_Jul
\item month\_of\_diagnosis\_Jun
\item month\_of\_diagnosis\_Mar
\item month\_of\_diagnosis\_May
\item month\_of\_diagnosis\_Nov
\item month\_of\_diagnosis\_Oct
\item month\_of\_diagnosis\_Sep
\item number\_of\_primaries
\item race\_ethnicity\_Amerian Indian, Aleutian, Alaskan Native or Eskimo
\item race\_ethnicity\_Asian Indian
\item race\_ethnicity\_Asian Indian or Pakistani
\item race\_ethnicity\_Black
\item race\_ethnicity\_Chamorran
\item race\_ethnicity\_Chinese
\item race\_ethnicity\_Fiji Islander
\item race\_ethnicity\_Filipino
\item race\_ethnicity\_Guamanian
\item race\_ethnicity\_Hawaiian
\item race\_ethnicity\_Hmong
\item race\_ethnicity\_Japanese
\item race\_ethnicity\_Kampuchean
\item race\_ethnicity\_Korean
\item race\_ethnicity\_Laotian
\item race\_ethnicity\_Melanesian
\item race\_ethnicity\_Micronesian
\item race\_ethnicity\_New Guinean
\item race\_ethnicity\_Other
\item race\_ethnicity\_Other Asian
\item race\_ethnicity\_Pacific Islander
\item race\_ethnicity\_Pakistani
\item race\_ethnicity\_Polynesian
\item race\_ethnicity\_Samoan
\item race\_ethnicity\_Thai
\item race\_ethnicity\_Tongan
\item race\_ethnicity\_Unknown
\item race\_ethnicity\_Vietnamese
\item race\_ethnicity\_White
\item seer\_historic\_stage\_a\_Distant
\item seer\_historic\_stage\_a\_In situ
\item seer\_historic\_stage\_a\_Localized
\item seer\_historic\_stage\_a\_Regional
\item seer\_historic\_stage\_a\_Unstaged
\item sex\_Female
\item spanish\_hispanic\_origin\_Cuban
\item spanish\_hispanic\_origin\_Dominican Republic
\item spanish\_hispanic\_origin\_Mexican
\item spanish\_hispanic\_origin\_Non-Spanish/Non-hispanic
\item spanish\_hispanic\_origin\_Other specified Spanish/Hispanic origin (excludes Dominican Repuclic)
\item spanish\_hispanic\_origin\_Puerto Rican
\item spanish\_hispanic\_origin\_South or Central American (except Brazil)
\item spanish\_hispanic\_origin\_Spanish surname only
\item spanish\_hispanic\_origin\_Spanish, NOS; Hispanic, NOS; Latino, NOS
\item spanish\_hispanic\_origin\_Uknown whether Spanish/Hispanic or not
\item year\_of\_birth
\item year\_of\_diagnosis
\item month
\end{itemize}

and 
\codewhite{newtarget} is the target variable, indicating whether or not the subject died in month given by the value of the \codewhite{month} variable.

\subsection{S6 Text}
\label{S6_Text}

{\bf Breast Cancer Feature Selection} The feature set used as input into both the Random Forest and Neural Network models, after the transformation described in section~(\ref{subsec:transformation}) is given below and also available in full detail in the file 
\codewhite{NewPatientBreastML.html}.

\begin{itemize}[noitemsep]
\item cs\_tumor\_size
\item elevation
\item grade\_moderately differentiated
\item grade\_poorly differentiated
\item grade\_ndifferentiated; anaplastic
\item grade\_well differentiated
\item histology\_recode\_broad\_groupings\_adenomas and adenocarcinomas
\item histology\_recode\_broad\_groupings\_adnexal and skin appendage neoplasms
\item histology\_recode\_broad\_groupings\_basal cell neoplasms
\item histology\_recode\_broad\_groupings\_complex epithelial neoplasms
\item histology\_recode\_broad\_groupings\_cystic, mucinous and serous neoplasms
\item histology\_recode\_broad\_groupings\_ductal and lobular neoplasms
\item histology\_recode\_broad\_groupings\_epithelial neoplasms, NOS
\item histology\_recode\_broad\_groupings\_nerve sheath tumors
\item histology\_recode\_broad\_groupings\_unspecified neoplasms
\item lat
\item laterality\_Bilateral involvement, lateral origin unknown; stated to be single primary
\item laterality\_Paired site, but no information concerning laterality; midline tumor
\item laterality\_Right: origin of primary
\item lng
\item marital\_stats\_at\_dx\_Divorced
\item marital\_stats\_at\_dx\_Married (inclding common law)
\item marital\_stats\_at\_dx\_Separated
\item marital\_stats\_at\_dx\_Single (never married)
\item marital\_stats\_at\_dx\_Unknown
\item marital\_stats\_at\_dx\_Unmarried or domestic partner
\item marital\_stats\_at\_dx\_Widowed
\item month\_of\_diagnosis\_Apr
\item month\_of\_diagnosis\_Aug
\item month\_of\_diagnosis\_Dec
\item month\_of\_diagnosis\_Feb
\item month\_of\_diagnosis\_Jan
\item month\_of\_diagnosis\_Jul
\item month\_of\_diagnosis\_Jun
\item month\_of\_diagnosis\_Mar
\item month\_of\_diagnosis\_May
\item month\_of\_diagnosis\_Nov
\item month\_of\_diagnosis\_Oct
\item month\_of\_diagnosis\_Sep
\item race\_ethnicity\_Amerian Indian, Aletian, Alaskan Native or Eskimo
\item race\_ethnicity\_Asian Indian
\item race\_ethnicity\_Black
\item race\_ethnicity\_Chinese
\item race\_ethnicity\_Japanese
\item race\_ethnicity\_Melanesian
\item race\_ethnicity\_Other
\item race\_ethnicity\_Other Asian
\item race\_ethnicity\_Pacific Islander
\item race\_ethnicity\_Thai
\item race\_ethnicity\_Unknown
\item race\_ethnicity\_Vietnamese
\item race\_ethnicity\_White
\item seer\_historic\_stage\_a\_Distant
\item seer\_historic\_stage\_a\_In sit
\item seer\_historic\_stage\_a\_Localized
\item seer\_historic\_stage\_a\_Unstaged
\item sex\_Female
\item spanish\_hispanic\_origin\_Cuban
\item spanish\_hispanic\_origin\_Mexican
\item spanish\_hispanic\_origin\_Non-Spanish/Non-hispanic
\item spanish\_hispanic\_origin\_Other specified Spanish/Hispanic origin (excldes Dominican Republic)
\item spanish\_hispanic\_origin\_Spanish surname only
\item spanish\_hispanic\_origin\_Spanish, NOS; Hispanic, NOS; Latino, NOS
\item year\_of\_birth
\item year\_of\_diagnosis
\item month
\end{itemize}

and 
\codewhite{newtarget} is the target variable, indicating whether or not the subject died in month given by the value of the \codewhite{month} variable.

\subsection{S7 Text}
\label{S7_Text}

{\bf Pseudocode for the Data Transformation}

\begin{verbatim}
def train(X, T, D)
    // X, T, D are the original dataset
    X' = []
    D' = []

    // the transformation
    for each index i in X:
        for t=1 to T[i]:
            new_D = (0 if t < T[i], else D[i])
            append new_D to D'
            new_X = (X[i], t)
            append new_X to X'

    return a decision tree trained on (X', D')

def pmf(h, X)
    // X is a single datapoint
    // returns an array A where A[i] = P(Y = i | X)
    A = []
    p_so_far = 1 // this is p(T >= t | X)
    for t = 1 to (the last month where h has any data):
        // h knows p(T = t | T >= t, X), we call this p_cur
        p_cur = h's prediction for (X, t)
        append (p_so_far * p_cur) to A
        p_so_far *= (1 - p_cur)

\end{verbatim}

\subsection{S8 Text}
\label{S8_Text}

{\bf Breast Random Forest Model Hyperparameters}
\begin{verbatim}
rf = RandomForestClassifier(n_estimators=20,min_samples_split=3,
                             max_depth = 15,
                            max_features = .8,
                             n_jobs=5,verbose=2,random_state=33)
\end{verbatim}

\subsection{S9 Text}
\label{S9_Text}
{\bf Colon Random Forest Model Hyperparameters}

\begin{verbatim}
rf = RandomForestClassifier(n_estimators=25,min_samples_split=3,
                             max_depth = 10,
                            max_features = .5,
                             n_jobs=5,verbose=2,random_state=3)
\end{verbatim}

\subsection*{S10 Text}
\label{S10_Text}
{\bf Lung Random Forest Model Hyperparameters}

\begin{verbatim}
rf = RandomForestClassifier(n_estimators=25,min_samples_split=3,
                             max_depth = 11,
                            max_features = .8,
                             n_jobs=5,verbose=2,random_state=3)
\end{verbatim}

\subsection{S11 Text}
\label{S11_Text}
{\bf Breast Neural Network Model Architecture}
The archictecture of the Keras multilayer perceptron neural network model 
trained on the breast cancer data is given explicitly below:

\begin{verbatim}
modelbreast = Sequential()
modelbreast.add(Dense(114, input_shape=(66,) ,init='normal'))
modelbreast.add(Activation('relu'))
modelbreast.add(Dropout(0.05))
modelbreast.add(Dense(50, init='normal'))
modelbreast.add(Activation('relu'))
modelbreast.add(Dropout(0.05))

modelbreast.add(Dense(36, init='normal'))
modelbreast.add(Activation('relu'))
modelbreast.add(Dropout(0.05))

modelbreast.add(Dense(2, init='normal'))
modelbreast.add(Activation('softmax'))

rms = RMSprop(lr=0.001)

modelbreast.compile(loss='binary_crossentropy', 
             optimizer=rms, class_mode="binary")

\end{verbatim}

and trained with a batch size of 1500 for 200 epochs.

\subsection{S12 Text}
\label{S12_Text}
{\bf Colon Cancer Neural Network Model Architecture}
The archictecture of the Keras multilayer perceptron neural network model 
trained on the colon cancer data is given explicitly below:

\begin{verbatim}


modelcolon = Sequential()
modelcolon.add(Dense(114, input_shape=(102,) ,init='normal'))
modelcolon.add(Activation('relu'))
modelcolon.add(Dropout(0.05))
modelcolon.add(Dense(50, init='normal'))
modelcolon.add(Activation('relu'))
modelcolon.add(Dropout(0.05))


modelcolon.add(Dense(35, init='normal'))
modelcolon.add(Activation('relu'))
modelcolon.add(Dropout(0.05))

modelcolon.add(Dense(2, init='normal'))
modelcolon.add(Activation('softmax'))

rms = RMSprop(lr=0.001)

modelcolon.compile(loss='binary_crossentropy',
          optimizer=rms, class_mode="binary")

\end{verbatim}

and trained with a batch size of 1500 for 200 epochs.

\subsection{S13 Text}
\label{S13_Text}
{\bf Lung Cancer Neural Network Model Architecture}

The archictecture of the Keras multilayer perceptron neural network model 
trained on the lung cancer data is given explicitly below:

\begin{verbatim}

modellung = Sequential()
modellung.add(Dense(114, input_shape=(114,) ,init='normal'))
modellung.add(Activation('relu'))
modellung.add(Dropout(0.1))
modellung.add(Dense(80, init='normal'))
modellung.add(Activation('relu'))
modellung.add(Dropout(0.1))
modellung.add(Dense(40, init='normal'))
modellung.add(Activation('relu'))
modellung.add(Dropout(0.1))


modellung.add(Dense(2, init='normal'))
modellung.add(Activation('softmax'))


rms = RMSprop(lr=0.001)

modellung.compile(loss='binary_crossentropy',
              optimizer=rms, class_mode="binary")

\end{verbatim}

and trained with a batch size of 2000 for 50 epochs.

\subsection{S1 Table}
\label{S1_Table}
{\bf Examples of four columns in the untransformed dataset.}
\begin{table}[H]
\begin{adjustwidth}{-2.25in}{0in} 
\begin{tabular}{lrrrr}
\toprule
{} &  \codewhite{cs\_tumor\_size} &  \codewhite{year\_of\_birth} &  \codewhite{survival\_months} &  \codewhite{vital\_status\_recode\_Dead} \\
\midrule
newindex &                &                &        &            \\
205      &             60 &           1951 &      3 &          1 \\
306      &            40  &         1950  &      3    &     0    \\
\bottomrule
\end{tabular}
\end{adjustwidth}
\end{table}

\subsection{S2 Table}
\label{S2_Table}
{\bf Example of four columns in the transformed dataset.}
\begin{table}[H]
\begin{adjustwidth}{-2.25in}{0in} 
\begin{tabular}{lrrrr}
\toprule
{} &  \codewhite{cs\_tumor\_size} &  \codewhite{year\_of\_birth} &  \codewhite{month} &  \codewhite{newtarget} \\
\midrule
newindex &                &                &        &            \\
205      &             60 &           1951 &      0 &          0 \\
205      &             60 &           1951 &      1 &          0 \\
205      &             60 &           1951 &      2 &          0 \\
205      &             60 &           1951 &      3 &          1 \\
306      &             40 &           1950 &      0 &          0 \\
306      &             40 &           1950 &      1 &          0 \\
306      &             40 &           1950 &      2 &          0  \\
306      &             40 &           1950 &      3 &          0  \\
\bottomrule
\end{tabular}
\end{adjustwidth}
\end{table}

\subsection{S3 Table}
\label{S3_Table}
{\bf AUC values for the Random Forest and Neural Networks model
binary classifiers derived from the full survival curve predictions; see text for details. The number of subjects that were used in the calculation of a given AUC score are given in parenthesis after the score.}
\begin{table}[!ht]
\begin{adjustwidth}{-2.25in}{0in} 
\begin{tabular}{lrrr}
\toprule
Model & 6 Months AUC & 12 Months AUC & 60 Months AUC \\ 
\midrule
Breast RF &  .846  (3035)     &     .885  (2797)         &  .844 (1392) \\ 
Breast NN &   .855 (3035)    &     .867  (2797)    &    .836  (1392) \\ 
Colon RF  &     .804 (5281)         &      .806 (5003)          &      .828   (3232)        \\ 
Colon NN   &     .797 (5281)         &          .804 (5003)        &   .841 (3232) \\ 
Lung RF    &      .772  (5019)             &        .796 (4860)              &   .874 (4143)  \\ 
Lung NN    &        .765  (5019)            &        .796  (4860)             &  .875 (4143)  \\
\bottomrule
\end{tabular}
\end{adjustwidth}
\end{table}

\subsection{S4 Table}
\label{S4_Table}
{\bf Percentage agreement for the Random Forest and Neural Network classifiers for 6, 12, and 60 month survival predictions on the test data for each cancer type.}
\begin{table}[!ht]
\begin{adjustwidth}{-2.25in}{0in} 
\begin{tabular}{lrrr}
\toprule
Cancer Type & $\%$ agreement 6 months & $\%$ agreement 12 months & $\%$ agreement 60 months \\ 
\midrule
Colon & .981 & .971 & .915 \\  
Breast & .994 & .984 & .938 \\  
Lung & .861 & .883 & .900 \\  
\bottomrule
\end{tabular}
\end{adjustwidth}
\end{table}

\subsection{S5 Table}
\label{S5_Table}
{\bf Correlations between the Random Forest and Neural Network classifiers for 6, 12, and 60 month survival predictions on the test data for each cancer type. The corresponding scatter plots are shown in Fig.~(\nameref{S2_Fig}).}
\begin{table}[!ht]
\begin{adjustwidth}{-2.25in}{0in} 
\begin{tabular}{lrrr}
\toprule
Cancer Type &  6 months &  12 months &   60 months \\ 
\midrule
Lung & .914 & .946 & .921 \\  
Colon & .900 & .923 & .935 \\  
Breast & .676 & .72 & .785 \\  
\bottomrule
\end{tabular}
\end{adjustwidth}
\end{table}

\subsection{S6 Table}
\label{S6_Table}
{\bf Example input data to the Colon Cancer neural network app \url{https://github.com/doolingdavid/colon-cancer-nn-errors.git}.}
\begin{table}[!ht]
\begin{adjustwidth}{-2.25in}{0in} 
\begin{tabular}{lr}
\toprule
  Variable  & Value \\ 
\midrule
  What is the tumor size (mm) & 300 \\  
  What is the patient's address? & boston massachusetts \\ 
  Grade & moderately differentiated \\  
  Histology & adenomas and adenocarcinomas \\ 
  Laterality & not a paired site \\  
 Martial Status at Dx & Single, never married \\  
 Month of Diagnosis & Jan \\  
 How many primaries & 1 \\  
  Race$\_$ethnicity & White \\  
  seer$\_$historic$\_$stage$\_$a  & Regional \\ 
  Gender & Male \\  
  spanish$\_$hispanic$\_$origin & Non-spanish/Non-hispanic \\ 
 Year of Birth & 1940 \\  
  Year of Diagnosis & 2010 \\
\bottomrule
\end{tabular}
\end{adjustwidth}
\end{table}

\subsection{S7 Table}
\label{S7_Table}
{\bf Example input data to the Lung Cancer random forest app \url{https://github.com/doolingdavid/lung-cancer-rf-errors.git}.}
\begin{table}[H]
\begin{adjustwidth}{-2.25in}{0in} 
\begin{tabular}{lr}
\toprule
  Variable  & Value \\ 
\midrule
  What is the tumor size (mm) & 500 \\  
  What is the patient's address? & newark new jersey \\ 
  Grade & well differentiated \\  
  Histology & acinar cell neoplasms \\ 
  Laterality & \makecell{bilateral involvement, lateral origin unknown; \\ stated to be single primary} \\  
 Martial Status at Dx & Married including common law \\  
 Month of Diagnosis & Jan \\  
 How many primaries & 1 \\  
  Race$\_$ethnicity & White \\  
  seer$\_$historic$\_$stage$\_$a  & Distant \\ 
  Gender & Female \\  
  spanish$\_$hispanic$\_$origin & Non-spanish/Non-hispanic \\ 
 Year of Birth & 1970 \\  
  Year of Diagnosis & 2011 \\
\bottomrule
\end{tabular}
\label{tab:lungmaritalstatus}
\end{adjustwidth}
\end{table}

\subsection{S8 Table}
\label{S8_Table}
{\bf Example of the transformation of \codewhite{STATE-COUNTY RECODE} to \codewhite{elevation}, \codewhite{lat}, and \codewhite{lng}.}
\begin{table}[H]
\begin{adjustwidth}{-2.25in}{0in} 
\begin{tabular}{llrrr}
\toprule
 STATE-COUNTY RECODE &               address &    elevation &        lat &         lng \\
\midrule
35001 &  Bernalillo+county+NM &  5207.579772 &  35.017785 & -106.629130 \\
35003 &      Catron+county+NM &  8089.242628 &  34.151517 & -108.427605 \\
35005 &      Chaves+county+NM &  3559.931671 &  33.475739 & -104.472330 \\
35006 &      Cibola+county+NM &  6443.415570 &  35.094756 & -107.858387 \\
35007 &      Colfax+county+NM &  6147.749089 &  36.579976 & -104.472330 \\
\bottomrule
\end{tabular}
\end{adjustwidth}
\end{table}

\subsection{S9 Table}
\label{S9_Table}
{\bf Filters applied to the Colon Cancer data.}
\begin{table}[H]
\begin{adjustwidth}{-2.25in}{0in} 
\begin{tabular}{lr}
\toprule
 Column &  Filter \\
\midrule
\codewhite{SEQUENCE NUMBER-CENTRAL} & \codewhite{$\neq$ "Unspecified"} \\
\codewhite{AGE AT DIAGNOSIS} & \codewhite{$\neq$ "Unknown age"} \\
\codewhite{BIRTHDATE-YEAR} & \codewhite{$\neq$ "Unknown year of birth"} \\
\codewhite{YEAR OF DIAGNOSIS} & \codewhite{$\geq 2004$} \\
\codewhite{SURVIVAL MONTHS FLAG} & \codewhite{= "1"}\\
\codewhite{CS TUMOR SIZE EXT/EVAL} & \codewhite{$\neq$ ""} \\
\codewhite{CS TUMOR SIZE} & \codewhite{$\neq 999$} \\
\codewhite{SEER RECORD NUMBER} & \codewhite{$= 1$} \\
\codewhite{PRIMARY SITE} & \codewhite{ $=$ "LARGE INTESTINE, (EXCL. APPENDIX)"} \\
\codewhite{SEQUENCE NUMBER-CENTRAL} & \codewhite{$=0$} \\
\bottomrule
\end{tabular}
\end{adjustwidth}
\end{table}

\subsection{S10 Table}
\label{S10_Table}
{\bf Filters applied to the Lung Cancer data.}
\begin{table}[H]
\begin{adjustwidth}{-2.25in}{0in} 
\begin{tabular}{lr}
\toprule
 Column &  Filter \\
\midrule
\codewhite{SEQUENCE NUMBER-CENTRAL} & \codewhite{$\neq$ "Unspecified"} \\
\codewhite{AGE AT DIAGNOSIS} & \codewhite{$\neq$ "Unknown age"} \\
\codewhite{BIRTHDATE-YEAR} & \codewhite{$\neq$ "Unknown year of birth"} \\
\codewhite{YEAR OF DIAGNOSIS} & \codewhite{$\geq 2004$} \\
\codewhite{SURVIVAL MONTHS FLAG} & \codewhite{= "1"}\\
\codewhite{CS TUMOR SIZE EXT/EVAL} & \codewhite{$\neq$ ""} \\
\codewhite{CS TUMOR SIZE} & \codewhite{$\neq 999$} \\
\codewhite{SEER RECORD NUMBER} & \codewhite{$= 1$} \\
\codewhite{PRIMARY SITE} & \codewhite{ $=$ "LUNG \& BRONCHUS"} \\
\codewhite{SEQUENCE NUMBER-CENTRAL} & \codewhite{$=0$} \\
\bottomrule
\end{tabular}
\end{adjustwidth}
\end{table}

\subsection{S11 Table}
\label{S11_Table}
{\bf Filters applied to the Breast Cancer data.}
\begin{table}[H]
\begin{adjustwidth}{-2.25in}{0in} 
\begin{tabular}{lr}
\toprule
 Column &  Filter \\
\midrule
\codewhite{SEQUENCE NUMBER-CENTRAL} & \codewhite{$\neq$ "Unspecified"} \\
\codewhite{AGE AT DIAGNOSIS} & \codewhite{$\neq$ "Unknown age"} \\
\codewhite{BIRTHDATE-YEAR} & \codewhite{$\neq$ "Unknown year of birth"} \\
\codewhite{YEAR OF DIAGNOSIS} & \codewhite{$\geq 2004$} \\
\codewhite{SURVIVAL MONTHS FLAG} & \codewhite{= "1"}\\
\codewhite{CS TUMOR SIZE EXT/EVAL} & \codewhite{$\neq$ " "} \\
\codewhite{CS TUMOR SIZE} & \codewhite{$\neq 999$} \\
\codewhite{SEER RECORD NUMBER} & \codewhite{$= 1$} \\
\codewhite{SEQUENCE NUMBER-CENTRAL} & \codewhite{$=0$} \\
\bottomrule
\end{tabular}
\end{adjustwidth}
\end{table}

\nolinenumbers

%
%
%




\bibliography{newbib}

\end{document}